\documentclass[a4paper]{article}
\usepackage{url}
\usepackage{hyperref}
\usepackage{appendix}
\usepackage{xcolor}
\usepackage{graphicx}
\usepackage{float}
\usepackage{multicol}
\usepackage{fancyhdr}
\usepackage{listings}
\usepackage{caption}
\usepackage{amsmath}
\usepackage{report_style}
\usepackage{siunitx}

\hypersetup{
    colorlinks=true,
    linkcolor=blue,
    urlcolor=blue
}

\usepackage{hyperref}
\usepackage[hyphenbreaks]{breakurl}

\DeclareCaptionFormat{codecaption}{#3} 
\lstdefinestyle{pythoncode}{
    language=Python,
    basicstyle=\ttfamily\footnotesize,
    numbers=left,
    numberstyle=\tiny,
    stepnumber=1,
    numbersep=5pt,
    xleftmargin=15pt,
    showspaces=false,
    showstringspaces=false,
    showtabs=false,
    framextopmargin=5pt,
    framexbottommargin=5pt,
    frame=tb,
    rulecolor=\color{black},
    tabsize=4,
    captionpos=b,
    breaklines=true,
    breakatwhitespace=true,
    title=\lstname,
    commentstyle=\color{white},
}

\title{Pricing European Options with Google AutoML, TensorFlow, and XGBoost}

\author{
Juan Esteban Berger \\ University of Notre Dame \\ jberger8@nd.edu \\
July 1, 2023
}

\begin{document}
\maketitle

\begin{abstract}
Researchers have been using Neural Networks and other related machine-learning techniques to price options since the early 1990s. After three decades of improvements in machine learning techniques, computational processing power, cloud computing, and data availability, this paper is able to provide a comparison of using Google Cloud's AutoML Regressor, TensorFlow Neural Networks, and XGBoost Gradient Boosting Decision Trees for pricing European Options. All three types of models were able to outperform the Black Scholes Model in terms of mean absolute error. These results showcase the potential of using historical data from an option's underlying asset for pricing European options, especially when using machine learning algorithms that learn complex patterns that traditional parametric models do not take into account.
\end{abstract}

\section{Introduction}

The XGBoost algorithm has been regarded as the gold standard for machine learning since its inception in 2014. Neural Networks have also showcased incredible abilities in learning extremely complicated patterns in data with large numbers of input variables. Most recently, however, machine learning practitioners have praised Google Cloud's AutoML models for their ease of use and incredible accuracy. This study hopes to discover if TensorFlow deep learning models, XGBoost gradient boosted decision trees, and Google Cloud's AutoML Regressor can outperform the Black Scholes model.

\section{Related Work}

In 2019, Stanford University students Yang and Ke \cite{ke2019option} studied long-short term memory neural networks and feed-forward neural networks (referred to as multilayer perceptrons in their paper) for the purpose of pricing options. These students used data sourced from OptionMetrics' IvyDB US dataset \cite{optionmetrics}, the same data source used for training the models in this paper. Yang and Ke decided not to use the implied volatility listed in the IvyDB US dataset. They instead used the past 20 lags in the underlying asset's closing price to train their long short term memory neural networks with the hopes that their models will be able to price options without using implied volatility as an input. However, unlike in Yang and Ke's paper, no recurrent neural networks were used in this study, but instead, the past 20 daily closing prices for the underlying asset were used as individual features (along with an options' strike price, current underlying price, risk-free rate, dividend yield, and whether the option was a call or a put) to train the feed-forward neural networks used in this study. Finally, Yang and Ke assessed the performance of their models according to the mean squared error and mean absolute percentage error (among other performance metrics). The results of Yang and Ke's study show how both standard feed-forward neural networks (i.e., multilayer perceptrons) and long short term memory neural networks are able to outperform the Black-Scholes Model in both mean squared error and mean absolute percentage error.

Furthermore, Neural Networks have been studied for the purposes of pricing options since the early 1990s. In 1993, Malliaris and Salchenberger \cite{malliaris1993beating} implemented Neural Networks that used an option's underlying price, strike price, time to expiration, implied volatility, risk-free rate, and past lags of both the option's price and the underlying asset's price to predict an option's current price. Even as early as 1993, the results of that study showed how Neural Networks could outperform the Black-Scholes model for both in-the-money and out-of-the-money options. Another interesting study was performed by le Roux and du Toit \cite{leRoux2001}. In this study, an option's underlying price, strike price, time to maturity, implied volatility, and risk-free rate are used to estimate the price of an option. This study showed that Neural Networks were able to emulate the Black-Scholes model with an accuracy of at least 99.5\% with a confidence interval of 96\%. These results are remarkable, considering the relatively simple neural network architectures and the lack of computing power in the early 1990s and 2000s.

More modern studies have been performed focused on pricing options with Neural Networks like those by Mitra in 2012 \cite{mitra2012option}, and by Can and Fadda in 2014 \cite{can2014nonparametric}. The study by Mitra used an option's underlying asset's price, an option's strike price, the time to maturity, the historical volatility of the underlying asset's price, and the risk-free rate to predict an option's price. It showed that neural networks could be used to improve theoretical option pricing approaches since they can learn features that are hard to incorporate into the classical approaches. Furthermore, Can and Fadda used a slightly different approach to pricing options by using an option's underlying price divided by its strike price as one of the features along with the option's underlying price, the time to maturity, and the risk-free rate to estimate the value of the option's price divided by the strike price. Once again, Neural Networks are shown to outperform the Black-Scholes model in terms of Mean Absolute Error. Finally, a highly thorough literature review was performed in 2020 by Ruf and Wang \cite{ruf2020neural}, where they summarized the methods and findings of using Neural Networks for Options Pricing and Hedging from the 1990s up until modern findings from 2019.
\section{Dataset}
The data for this study was sourced from the OptionMetrics' IvyDB US dataset. OptionMetrics is a financial research and consulting firm that provides historical options data and analytics on global exchange-traded options. It is a subsidiary of Wharton Research Data Services. The IvyDB US dataset includes many tables with the historical market, options, and securities data ranging from 1990 to 2021 (as of this writing). The final dataset used to train the models consisted of 10,750,886 observations. The target variable was the midpoint price (which was calculated as the average between the bid and ask price for a given function. The feature variables were the option's strike price, implied volatility, the zero coupon rate, the index dividend yields, the option type (either call or put), the time to maturity in years, and the underlying assets' current price. The underlying asset's past 20 days' closing prices were also included as 20 additional features. The dataset was filtered so only European Options had indexes as underlying assets. Furthermore, only options with midpoint prices less than 100,000 were selected with the goal of eliminating extreme outliers.
Furthermore, big data technologies had to be used for querying and cleaning the data since the original dataset was over 500GB large. To efficiently query the data, BigQuery was utilized, and the data was cleaned using Google SQL. Finally, the data was split into training, validation, and testing datasets with approximate splits of 98\% of the data being used for training, 1\% of the data being used for validation, and 1\% being used for testing.

\section{Models}
Six models were implemented with Python for pricing options. These models were the Black-Scholes Model, a three-Layer Feed-Forward Neural Network, a five-Layer feed-forward Neural Network, a gradient-boosted decision tree with a max depth of five, a Gradient Boosted Decision Tree with a Max Depth of ten, and a Google Cloud AutoML Regressor. The feed-forward neural networks were implemented with the Tensorflow Framework and trained on Google Colab's TPU's (which had eight devices available). The gradient-boosted decision tree models were implemented using the XGBoost Framework and trained on Google Colab's A100 GPU. The Google AutoML Model was trained in Google Cloud's Vertex AI platform.

\subsection{Black-Scholes Model}
The options used for this study were all European options and can be priced according to the Black-Scholes Model:\\
$$
C = Se^{-qT}N(d_1)-Ke^{-rT}N(d_2)
$$
$$
P = Ke^{-rT}N(-d_2)-Se^{-qT}N(-d_1)
$$
where
$$
d_1 = [\ln(S/K) + (r - q + \tfrac{1}{2}\sigma^2)T] / \sigma\sqrt{T},
$$
$$
d_2 = d_1 - \sigma\sqrt{T}.
$$
\\
In the Black-Scholes model, $C$ is the price of a call option, $P$ is the price of a put option, $S$ is the current underlying security price, $K$ is the strike price of the option, $T$ is the time in years remaining to an options' expiration date, $r$ is the continuously-compounded interest rate, $q$ is the continuously-compounded annualized dividend yield, and $sigma$ is the implied volatility. This is the only model in this study that used implied volatility as one of the inputs and the only model in this study that does not use any past lags of the underlying security's closing price as inputs.

\subsection{Feed-Forward Neural Network Models}
This study implemented two different Feed-Forward Neural Networks. The first one was a three-layer Feed-Forward Neural Network, and the second was a three-layer Feed-Forward Neural Network. All the neural network models were trained using the Keras framework with the TensorFlow backend. The models were trained on a dataset that has been preprocessed by separating the Bid-Ask Midpoint Price column from the rest of the dataset and splitting the data into training, validation, and test sets. Unlike the Black-Scholes Model, the implied volatility column was dropped, and the past 20 lags were added as individual feature variables with the hope that the Neural Networks would be able to learn the necessary features to predict the option's midpoint price. The other feature variables were the underlying securities' price, the option's strike price, the time to maturity, the risk-free rate, the underlying index's dividend yield, and a binary variable indicating whether an option is a call or a put. The training processes use the Adam optimizer with an adaptive learning rate that starts at 0.01 and decreases by a factor of 0.1 every 10 epochs that it doesn't see an increase in performance until it reaches a learning rate of \(1 \times 10^{-6}\), and an early stopping callback if there isn't an improvement in performance in the last 150 epochs. The models are trained on Google Cloud's TPUs (Tensor Processing Units) with eight available devices for accelerated training.

The three-layer Feed-Forward Neural Network's first hidden layer has 256 neurons with the rectified linear unit (ReLU) activation function. The second hidden layer has 128 neurons with ReLU activation function, and the output layer has a single neuron with the linear activation function. The code for the three-layer feed-forward neural network written is shown below:
\begin{lstlisting}[style=pythoncode, caption=3 Layer Feed-Forward Neural Network, label=lst:model_2]
model = Sequential([
  Dense(256, input_dim=X_train.shape[1], activation='relu'),
  Dense(128, activation='relu'),
  Dense(1, activation='linear')
])                  
\end{lstlisting}
The five-layer Feed-Forward Neural Network's first hidden layer has 256 neurons with the rectified linear unit (ReLU) activation function. The second hidden layer has 128 neurons with ReLU activation function. The third hidden layer has 64 neurons with ReLU activation function. The fourth layer has 32 neurons with ReLU activation function, and the output layer has a single neuron with the linear activation function. The for five-layer feed-forward neural network written in Python can be seen below:
\begin{lstlisting}[style=pythoncode, caption=5 Layer Feed-Forward Neural Network, label=lst:model_4]
model = Sequential([
  Dense(256, input_dim=X_train.shape[1], activation='relu'),
  Dense(128, activation='relu'),
  Dense(64, activation='relu'),
  Dense(32, activation='relu'),
  Dense(1, activation='linear')
])
\end{lstlisting}

\subsection{Gradient Boosted Decision Tree Models}

The next two models were implemented with XGBoost, a gradient boosting algorithm, for predicting the Bid-Ask Midpoint Price of the options in the dataset. The models were trained on a dataset that has been preprocessed by separating the Bid-Ask Midpoint Price column from the rest of the dataset and splitting the data into training, validation, and test sets. Just like the Neural Network models, the implied volatility column was dropped, and the past 20 lags were added as individual feature variables with the hope that the Neural Networks would be able to learn the necessary features to predict the option's midpoint price. The other feature variables were the underlying securities' price, the option's strike price, the time to maturity, the risk-free rate, the underlying index's dividend yield, and a binary variable indicating whether an option is a call or a put.

In order to use an adaptive learning rate, the custom learning rate function was implemented so that the learning rate would decrease as the boosting rounds would advance for the optimization steps to become gradually smaller as shown below:

\begin{lstlisting}[style=pythoncode, caption=Customized Learning Rate Scheduler, label=lst:eta_decay]
def eta_decay(iteration):
    max_iter = 100000
    x = iteration + 1
    eta_base = 0.5
    eta_min = 0.2
    eta_decay = eta_min + (eta_base - eta_min) * np.exp(-(x/8)**2 / max_iter)
    return eta_decay
\end{lstlisting}

The XGBoost Framework was used to train two models, one with a maximum depth of five and the other with a maximum depth XGBoost Model with a maximum depth of ten. The code used to train the XGBoost Model with a maximum depth of five written in Python can be seen below:

\begin{lstlisting}[style=pythoncode, caption=XGBoost Model with Max Depth of 5, label=lst:model_5]
max_iter = 40000
eta_decay = np.array(
    [eta_decay(iteration) for iteration in range(max_iter)])
            
PARAMS = {
  'booster': 'gbtree',
  'eval_metric': 'mae',
  'max_depth': 5,
  'tree_method': 'gpu_hist'
}

evals_result = {
    'train': dtrain, 'validation': dval}

progress1 = dict()

model = xgb.train(
  maximize=True,
  params=PARAMS,
  dtrain=dtrain,
  num_boost_round=max_iter,
  early_stopping_rounds=max_iter,
  evals=[
    (dtrain, 'train'),(dtest, 'test')],
  evals_result=progress1,
  verbose_eval=1,
  callbacks=[
    xgb.callback.LearningRateScheduler(
    lambda iteration: eta_decay[iteration])])
\end{lstlisting}
The code used to train the XGBoost Model with a maximum depth of five written in Python is exactly the same as the code used to train the XGBoost Model with a maximum depth of ten except for the fact that the max depth was set to ten as shown below:
\begin{lstlisting}[style=pythoncode, caption = XGBoost Parameters for Max Depth of 10, label=lst:model_6]
PARAMS = {
  'booster': 'gbtree',
  'eval_metric': 'mae',
  'max_depth': 10,
  'tree_method': 'gpu_hist'
}
\end{lstlisting}

\subsection{Google Cloud AutoML Regressor}

The final model in this study was trained on Google Cloud's Vertex AI Platform. The model chosen was the AutoML Regressor. The model was given a budget of 72 node hours, but early stopping was implemented, and the model was only trained for two days and 27 minutes. The model's objective was tabular regression, and it was optimized for minimizing mean absolute error.

The models in this study were trained on the A100 GPUs provided by Google Colab. The use of GPUs significantly accelerated the training process and allowed for experimentation with more complex models. The use of these GPUs allowed for faster model training and tuning, enabling the exploration of a larger range of model architectures and hyperparameters.

\section{Results}
The testing results for all the models are summarized in the table below.

\begin{table}[htbp]
\centering
\caption{Models MAE}
\label{table:modelcomparison_mae_filt}
\begin{tabular}{cccc}
\hline
Model                   & MAE & MAPE & Training (s)\\
\hline
XGBoost 10     & 0.8093             & 42.23                   & 1917  \\
AutoML            & 1.0248             & 42.73                   & 174420 \\
XGBoost 5      & 1.6362             & 187.02                   & 971  \\
5 Layer FFNN      & 4.6374             & 243.90                   & 3288  \\
3 Layer FFNN      & 8.8075             & 323.77                   & 3066  \\
Black Scholes            & 8.0082             & 63.88                    & NA  \\
\hline
\end{tabular}
\end{table}

All of the models in this study were able to outperform the Black-Scholes model in terms of mean absolute error, which was the metric on which the trainable models were trained to minimize. The most accurate model was the XGBoost model with a max depth of ten. This model was ten times more accurate than the Black-Scholes model. Surprisingly, only two models were able to surpass the Black-Scholes model in terms of mean absolute percentage error and those were the XGBoost model with a max depth of ten and the Google AutoML Regressor. When comparing the two best-performing models, it is important to note that the AutoML Regressor took over two days to complete its training (with early-stopping enabled) while the XGBoost model with a max depth of ten was trained in a little over 30 minutes. Nevertheless, the AutoML regressor takes almost no expertise to train since all of the hyper-parameter tuning is done automatically. In order for the XGBoost model to beat the AutoML Regressor domain expertise is required as showcased by the use of custom learning-rate schedulers. It is important to note that the range of option prices used in this study range from 0.01 to 100,000. Extreme values, especially from out-of-the-money options, may lead to a skewed measure of mean absolute percentage error. Furthermore, all the trainable models were trained to optimize mean absolute error, not mean absolute percentage error.

\begin{figure}[H]
    \centering
    \includegraphics[width=\linewidth]{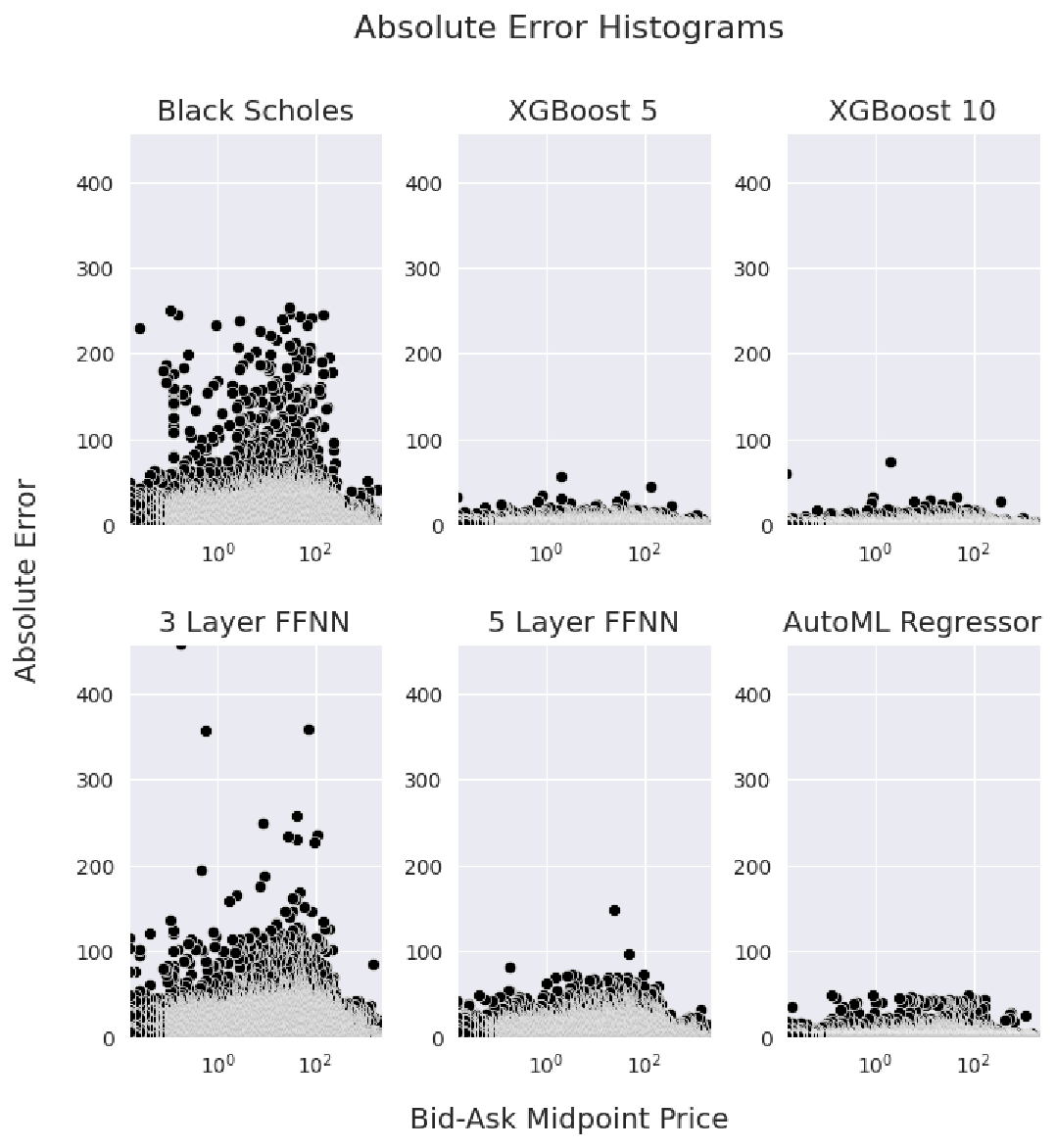}
    \caption{Absolute Error vs. Bid-Ask Midpoint Price}
    \label{fig:MAE}
  \end{figure}

\begin{figure}[H]
    \centering
    \includegraphics[width=\linewidth]{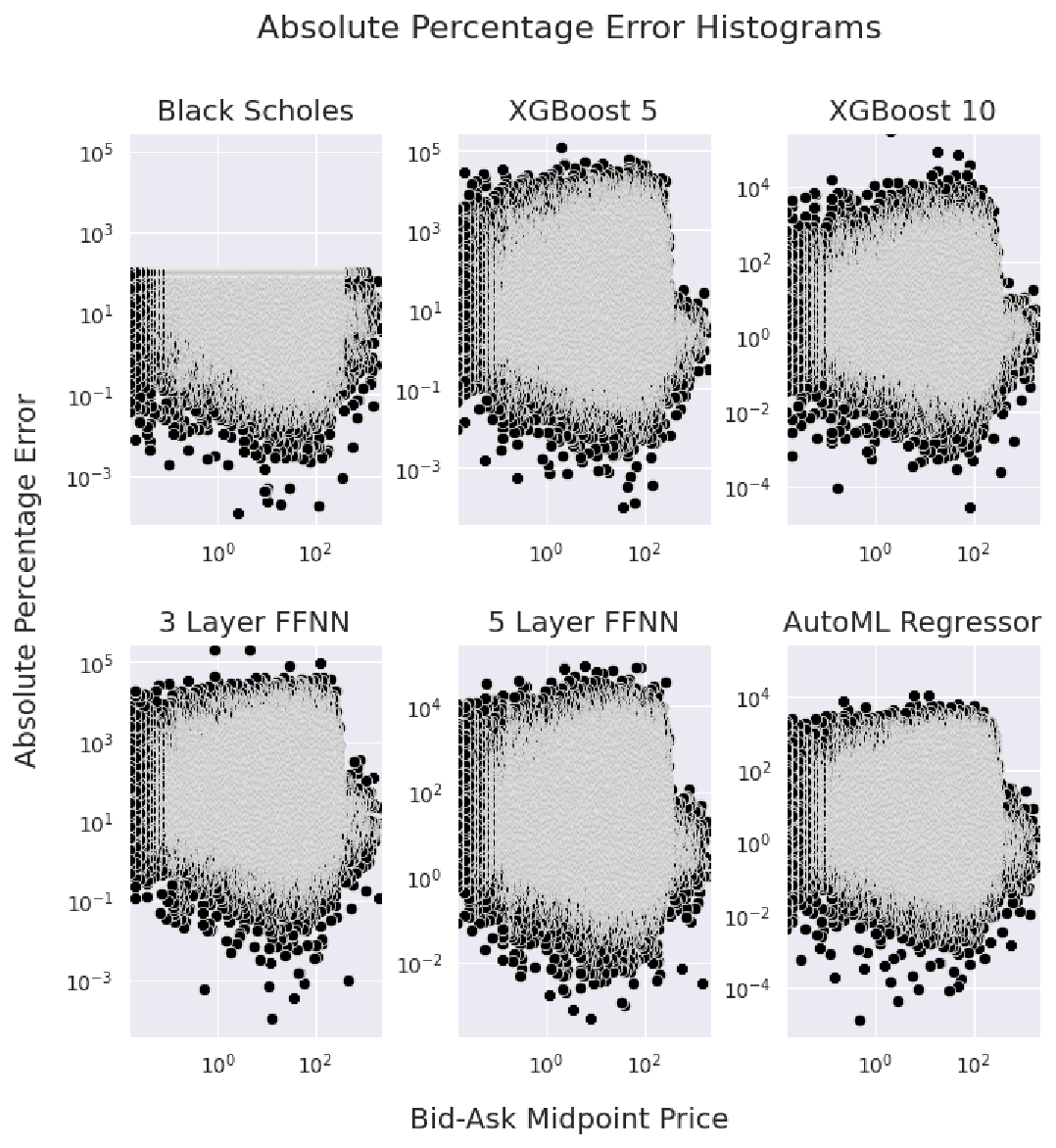}
    \caption{Absolute Percentage Error vs. Bid-Ask Midpoint Price}
    \label{fig:MAPE}
  \end{figure}

Given that the XGBoost models were able to outperform both TensorFlow and Google's AutoML Regressor, it makes sense why so many machine learning engineers and data scientists use this as their model of choice. All of the machine learning models were able to outperform the Black-Scholes model, but it is clear that the best performance in this study was achieved by the XGBoost models. The XGBoost with a max depth of ten had the lowest mean absolute error, mean absolute percentage error, and it was trained in a fraction of the time than its closest competitor (Google's Auto ML Regressor). It is also important to note that none of the machine learning models were given implied volatility as a feature like the Black-Scholes model. Instead, all the models learned all the necessary features on their own from the other feature variables and the past 20 lags of the underlying securities' closing prices. With high volumes of data and computing resources becoming highly available, using machine learning and deep learning methods for options pricing becomes a viable option for pricing securities. In further studies, it would be interesting to compare machine learning and deep learning methodologies with other option pricing methods such as Monte Carlo Simulations or the Binomial Asset Pricing Model in order to see if the Machine Learning models are able to outperform those pricing methodologies as well. Nevertheless, the results of this study seem to be consistent with the results of other related studies in that machine learning methods are able to outperform the Black-Scholes model, especially in the models created using XGBoost.

\section{Conclusion}
Given that the XGBoost models were able to outperform both TensorFlow and Google's AutoML Regressor, it makes sense why so many machine learning engineers and data scientists use this as their model of choice. All of the machine learning models were able to outperform the Black-Scholes model, but it is clear that the best performance in this study was achieved by the XGBoost models. The XGBoost with a max depth of ten had the lowest mean absolute error, mean absolute percentage error, and it was trained in a fraction of the time than its closest competitor (Google's Auto ML Regressor). It is also important to note that none of the machine learning models were given implied volatility as a feature like the Black-Scholes model. Instead, all the models learned all the necessary features on their own from the other feature variables and the past 20 lags of the underlying securities' closing prices. With high volumes of data and computing resources becoming highly available, using machine learning and deep learning methods for options pricing becomes a viable option for pricing securities. In further studies, it would be interesting to compare machine learning and deep learning methodologies with other option pricing methods such as Monte Carlo Simulations or the Binomial Asset Pricing Model in order to see if the Machine Learning models are able to outperform those pricing methodologies as well. Nevertheless, the results of this study seem to be consistent with the results of other related studies in that machine learning methods are able to outperform the Black-Scholes model, especially in the models created using XGBoost.

\section*{Acknowledgments}

This research project contains components for three graduate-level courses at the University of Notre Dame, which are ACMS 80695 - Master's Research Project taught by Prof. Guosheng Fu, CSE 60868 - Neural Networks taught by Prof. Adam Czajka, and ACMS 60890 - Statistical Foundations of Data Science taught by Prof. Xiufan Yu.  Special thanks are given to all these professors for their teaching and support during the Spring Semester of 2023.

The complete GitHub repository for the code in this project can be accessed through the following URL: \url{https://github.com/juan-esteban-berger/Options_Pricing_AutoML_TensorFlow_XGBoost/}. The most accurate model can be accessed through Hugging Face: \url{https://huggingface.co/juan-esteban-berger/XGBoost_European_Options_Pricing_MD_10}.


\onecolumn 

\appendix
\section{\label{sec:appendix}Appendix}

\begin{table}[htbp]
\centering
\caption{Summary Statistics for Financial Data}
\label{table:financialdata}
\begin{tabular}{ccccccc}
\hline
Statistic & Midpoint & Strike Price & Imp. Volatility & Zero Coupon Rate & Div. Yield & Time (years)\\
\hline
count & \num{10750890} & \num{10750890} & \num{10750890} & \num{10750890} & \num{10750890} & \num{10750890}\\
mean & \num{128.88} & \num{1369.92} & \num{0.51} & \num{2.90} & \num{2.36} & \num{0.59}\\
std & \num{306.93} & \num{2406.36} & \num{0.44} & \num{1.997617} & \num{2.04} & \num{0.53}\\
min & \num{0.015} & \num{5} & \num{0.0112} & \num{0.2935} & \num{0.000107} & \num{0.00274}\\
25\% & \num{6.5} & \num{275} & \num{0.2354} & \num{1.29} & \num{1.27} & \num{0.2163}\\
50\% & \num{32} & \num{620} & \num{0.3669} & \num{2.21} & \num{1.88} & \num{0.5202}\\
75\% & \num{112.1} & \num{1495} & \num{0.6167} & \num{4.58} & \num{2.54} & \num{0.7392}\\
max & \num{13267} & \num{21500} & \num{2.99999} & \num{7.63} & \num{147.35} & \num{4.7967}\\
\hline
\end{tabular}
\end{table}

\begin{multicols}{2}
  \begin{figure}[H]
    \centering
    \includegraphics[width=\linewidth]{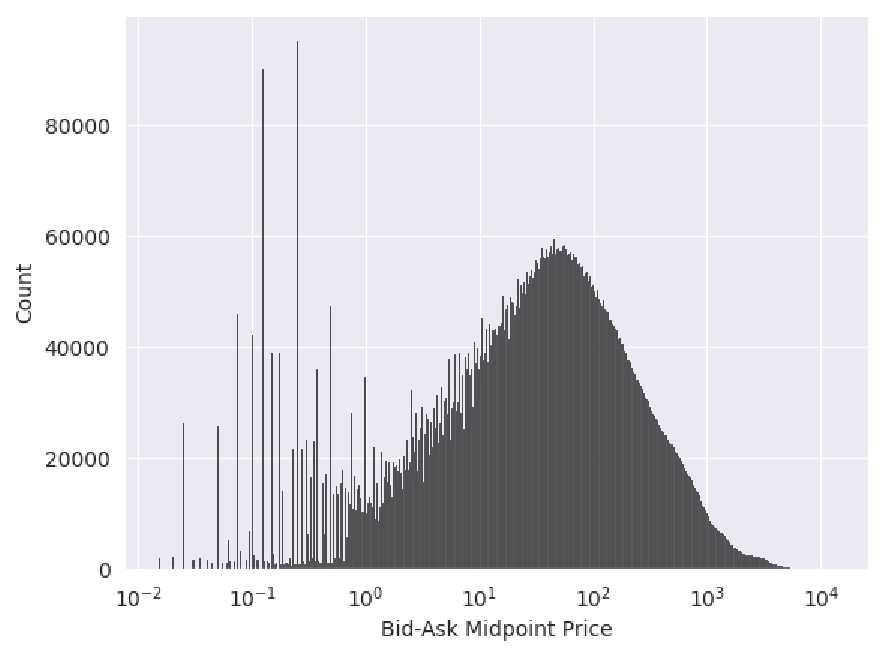}
    \caption{Distribution of Bid-Ask Midpoint Prices}
    \label{fig:01_midpoint_dist}
  \end{figure}
  
  \begin{figure}[H]
    \centering
    \includegraphics[width=\linewidth]{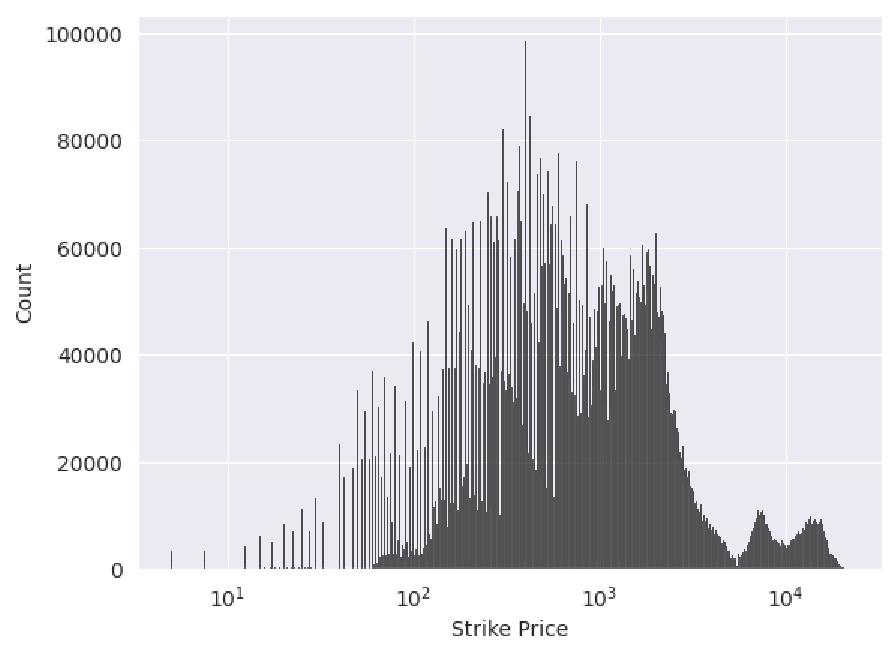}
    \caption{Distribution of Strike Prices}
    \label{fig:02_strike_price}
  \end{figure}
\end{multicols}

\begin{multicols}{2}
  \begin{figure}[H]
    \centering
    \includegraphics[width=\linewidth]{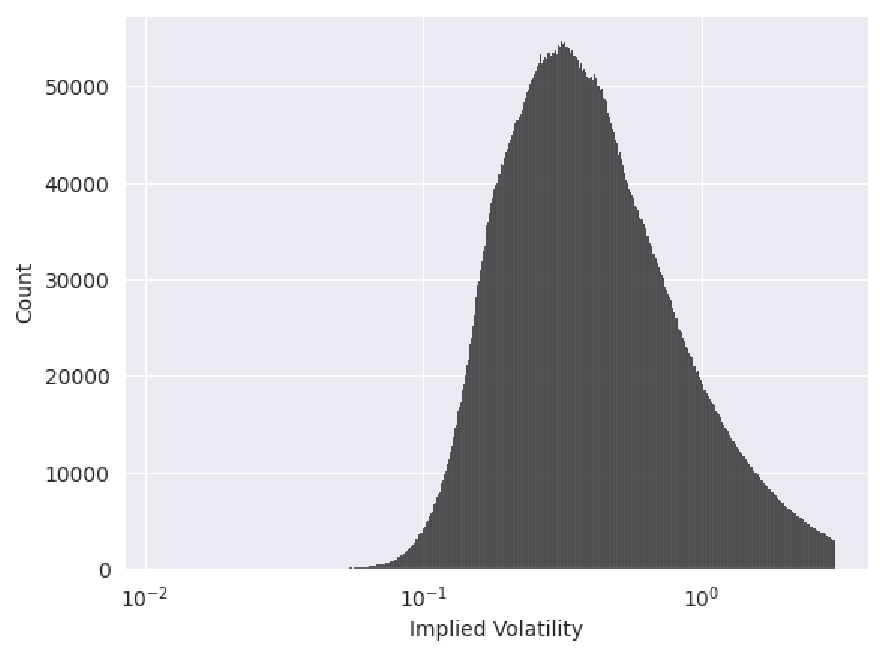}
    \caption{Distribution of Implied Volatilities}
    \label{fig:03_imp_vol_dist}
  \end{figure}
  
  \begin{figure}[H]
    \centering
    \includegraphics[width=\linewidth]{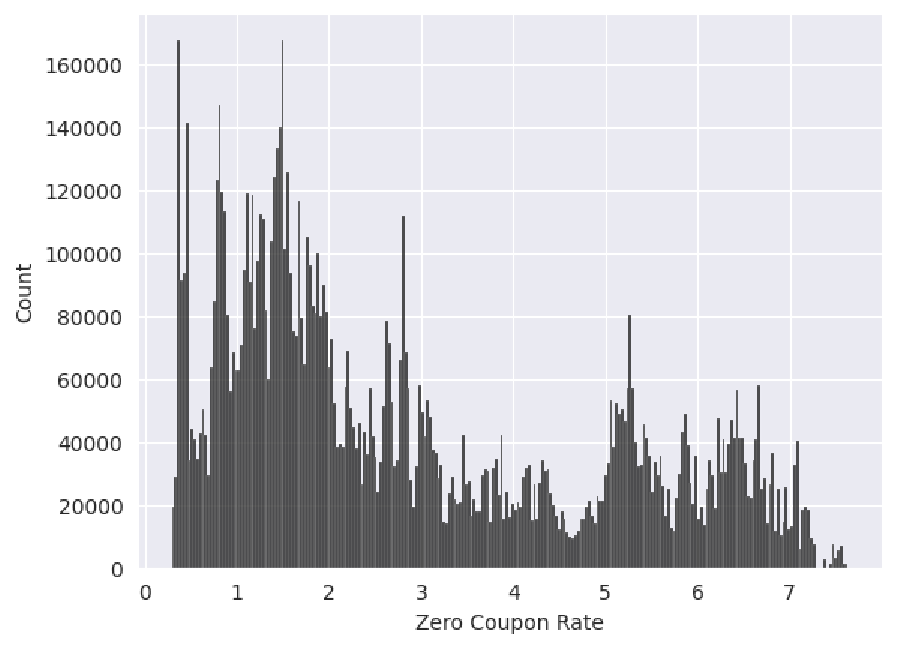}
    \caption{Distribution of Zero Coupon Rates}
    \label{fig:04_zero_coupon_dist}
  \end{figure}
\end{multicols}

\pagebreak

\begin{multicols}{2}
  \begin{figure}[H]
    \centering
    \includegraphics[width=\linewidth]{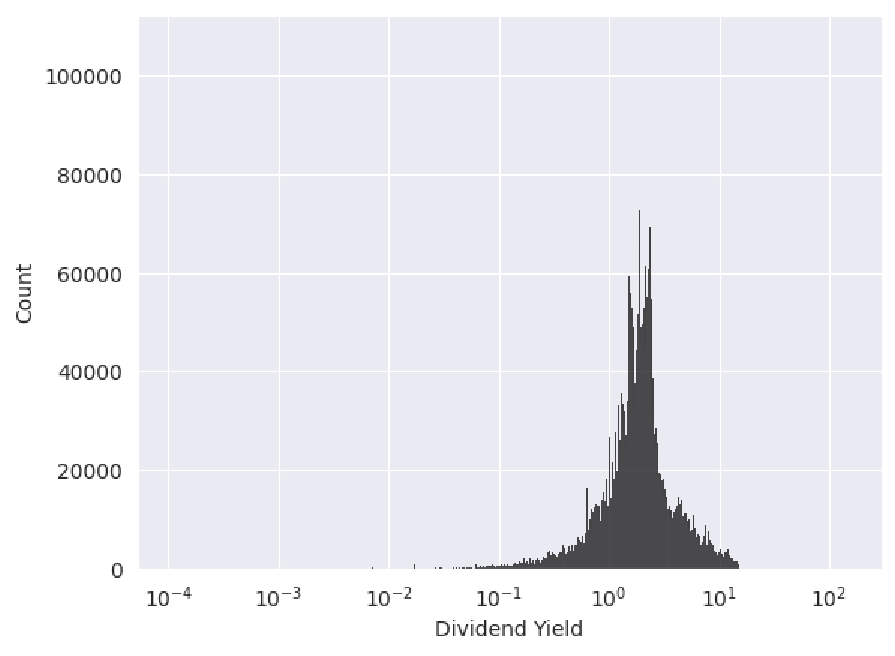}
    \caption{Distribution of Index Dividend Yields}
    \label{fig:05_div_yield_dist}
  \end{figure}
  
  \begin{figure}[H]
    \centering
    \includegraphics[width=\linewidth]{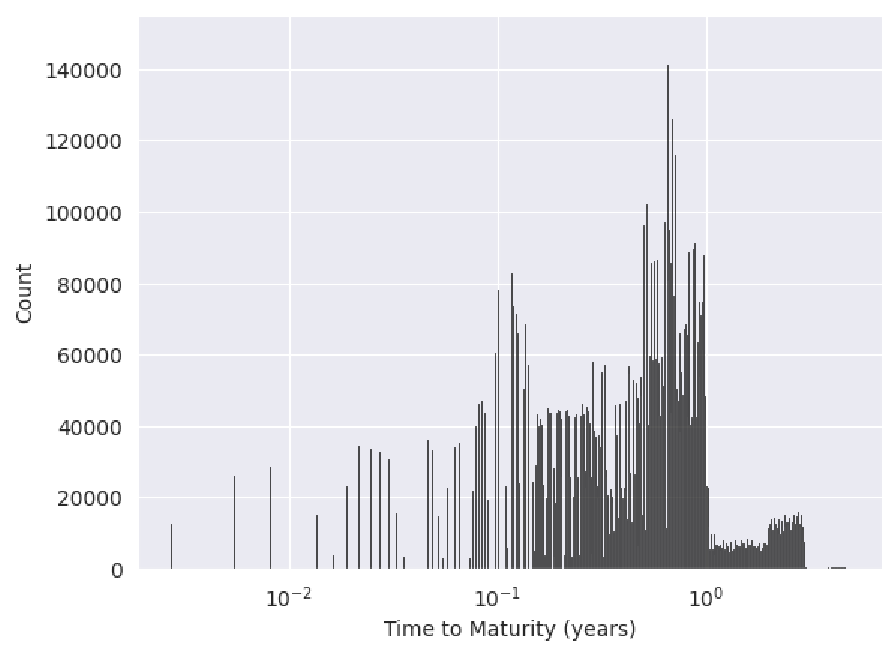}
    \caption{Distribution of Times to Maturity (years)}
    \label{fig:06_time_dist}
  \end{figure}
\end{multicols}

\begin{multicols}{2}
  \begin{figure}[H]
    \centering
    \includegraphics[width=\linewidth]{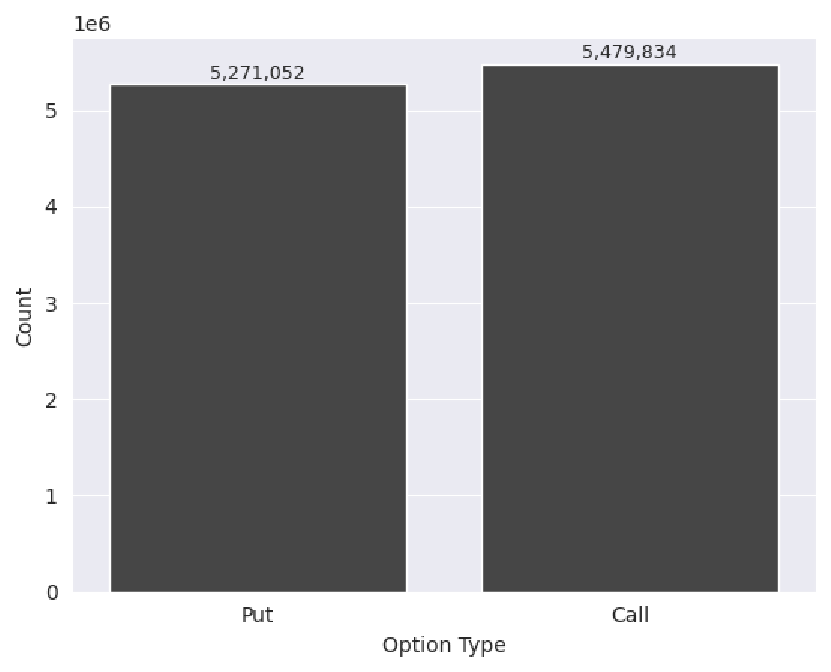}
    \caption{Distribution of Calls and Puts}
    \label{fig:07_opt_type_dist}
  \end{figure}
\end{multicols}

\end{document}